\documentclass[journal=jacsat,manuscript=article]{achemso}

\usepackage{chemformula} 
\usepackage[T1]{fontenc} 
\usepackage{amsthm,amsmath,amssymb}
\usepackage{mathrsfs,gensymb}
\usepackage{bm}
\usepackage{verbatim}
\usepackage{lineno}

\author{Jiawei Peng}
\affiliation[First University]
{School of Chemistry, South China Normal University, Guangzhou 510006, China}
\alsoaffiliation[Second University]
{MOE Key Laboratory of Environmental Theoretical Chemistry, South China Normal University, Guangzhou 510006, China}
\alsoaffiliation[Third University]
{SCNU Environmental Research Institute, Guangdong Provincial Key Laboratory of Chemical Pollution and Environmental Safety, School of Environment, South China Normal University, Guangzhou 510006, China}
\author{Deping Hu}
\affiliation[First University]
{Center for Advanced Materials Research, Beijing Normal University, Zhuhai 519087, China}
\author{Hong Liu}
\affiliation[First University]
{MOE Key Laboratory of Environmental Theoretical Chemistry, South China Normal University, Guangzhou 510006, China}
\alsoaffiliation[Second University]
{SCNU Environmental Research Institute, Guangdong Provincial Key Laboratory of Chemical Pollution and Environmental Safety, School of Environment, South China Normal University, Guangzhou 510006, China}
\email{hongliu@m.scnu.edu.cn}
\author{Qiang Shi}
\affiliation[First University]
{Beijing National Laboratory for Molecular Sciences, State Key Laboratory for Structural Chemistry of Unstable and Stable Species, Institute of Chemistry, Chinese Academy of Sciences, No. 2 North 1st Street, Zhongguancun, Beijing 100190, China}
\author{Peng Bao}
\affiliation[First University]
{Beijing National Laboratory for Molecular Sciences, State Key Laboratory for Structural Chemistry of Unstable and Stable Species, Institute of Chemistry, Chinese Academy of Sciences, No. 2 North 1st Street, Zhongguancun, Beijing 100190, China}
\author{Zhenggang Lan}
\affiliation[First University]
{MOE Key Laboratory of Environmental Theoretical Chemistry, South China Normal University, Guangzhou 510006, China}
\alsoaffiliation[Second University]
{SCNU Environmental Research Institute, Guangdong Provincial Key Laboratory of Chemical Pollution and Environmental Safety, School of Environment, South China Normal University, Guangzhou 510006, China}
\email{zhenggang.lan@m.scnu.edu.cn; zhenggang.lan@gmail.com.}
                                                                                               
\title[An \textsf{achemso} demo]                                                               
{Studies of Nonadiabatic Dynamics in the Singlet Fission Processes of Pentacene Dimer via Tensor Train Decomposition Method}                                            
\abbreviations{IR,NMR,UV}                                                                      
                                                                                               
\begin{document}
\newpage
\begin{abstract}
Singlet fission (SF) is a very significant photophysical phenomenon and possesses potential applications. In this work, we try to give the rather detailed theoretical investigation of the SF process in the stacked polyacene dimer by combining the high-level quantum chemistry calculations, and the quantum dynamics simulations based on the tensor train decomposition method. Starting from the construction of the linear vibronic coupling model, we explore the pure electronic dynamics and the vibronic dynamics in the SF processes.  The role of vibrational modes in nonadiabatic dynamics is addressed. The results show that the super-exchange mechanism mediated by the charge-transfer state is found in both pure electronic dynamics and the nonadiabatic dynamics. Particularly, the vibrational modes with the frequency resonance with the adiabatic energy gap play very import roles in the SF dynamics. This work not only provides a deep and detailed understanding of the SF process, but also verifies the efficiency of the tensor train decomposition method that can serve as the reference dynamics method to explore the dynamics behaviors of complex systems.
\end{abstract}

\newpage                                                                              
\section{I. Introduction}   
Singlet fission (SF) is a very significant photophysical phenomenon in a group of organic photovoltatic materials, such as covalent dimers (tetraene\cite{muller_evidence_2006}, pentacene\cite{zirzlmeier_singlet_2015} and diphenylisobenzofuran,\cite{schrauben_excitation_2016} etc.\cite{varnavski_high_2015,margulies_enabling_2016,zeng_design_2016}), polyenes and carotenoids,\cite{polivka_dark_2009} conjugated polymers,\cite{barford_electronic_2013} and molecular solids (thin film or crystal),\cite{barford_electronic_2013} due to its huge potential to boost the quantum yield of solar energy conversation.\cite{smith_singlet_2010,smith_recent_2013,casanova_theoretical_2018,miyata_triplet_2019,dimitriev_dynamics_2022}
In the SF process, one singlet exciton is split into a pair of the coupled triplet excitons, and then be converted into two noninterating triplet states. 
The SF process is spin-allowed since the two produced triplets are coupled into an overall singlet state. This indicates that the SF reaction may take place in the ultrafast timedomain from the femtosecond to picosecond timescale, which is usually not affected by the nanosecond-scale radiative decay.\cite{smith_singlet_2010,smith_recent_2013,casanova_theoretical_2018,miyata_triplet_2019,dimitriev_dynamics_2022}
In principle, if some organic photovoltatic materials with efficient SF reactions are properly designed, it is possible to go beyond the Shockley-Queisser limit,\cite{shockley_detailed_1961} largely improve the photoelectric conversion efficiency and  promote the commercial application of organic photovoltatic devices.\cite{smith_singlet_2010,zeng_seeking_2014} To achieve this goal, the exploration of the intrinsic nature behind the SF process becomes essential.



In the past decade, extensive experimental and theoretical effects were devoted to this research area.\cite{lee_influence_1977,sebastian_charge_1981,rei_vilar_spectroscopy_1983,jundt_exciton_1995,troisi_band_2005,marciniak_ultrafast_2007,marciniak_ultrafast_2009,johnson_high_2010,rao_exciton_2010,wang_high-yield_2010,kuhlman_interpretation_2010,zimmerman_singlet_2010,burdett_dependence_2011,chan_observing_2011,wilson_ultrafast_2011,zimmerman_mechanism_2011,rao_photophysics_2011,yamagata_nature_2011,chan_energy_2012,chan_harvesting_2012,burdett_quantum_2012,roberts_efficient_2012,ma_singlet_2012,wang_singlet_2012,wilson_temperature-independent_2013,tayebjee_exciton_2013,lee_singlet_2013-1,nelson_exceeding_2013,wen_singlet_2013,mastron_aqueous_2013,wong_revealing_2013,walker_singlet_2013,lee_singlet_2013,zimmerman_correlated_2013,sharifzadeh_low-energy_2013,zhang_nonlinear_2014,schrauben_mechanism_2014,kolata_molecular_2014,busby_multiphonon_2014,ambrosio_singlet_2014,yost_transferable_2014,casanova_electronic_2014,musser_evidence_2015,varnavski_high_2015,busby_design_2015,tamura_first-principles_2015,monahan_charge_2015,aryanpour_theory_2015,coto_low-lying_2015,piland_how_2015,herz2015unveiling,bakulin_vibrational_2015,zirzlmeier_solution-based_2016,basel_unified_2017,schroder_multi-dimensional_2017,fujihashi_effect_2017,monahan2017dynamics,breen2017triplet,reddy_quantum_2019,schroeder_tensor_2019,casillas_intermolecular_2019,duan_intermolecular_2020,smyser_singlet_2020,han_evaluation_2020,mardazad_quantum_2021,zhao_minimal-active-space_2021,bai_plenty_2022,claudino_modeling_2023,teichen_microscopic_2012,beljonne_charge-transfer_2013,berkelbach_microscopic_2013-1,berkelbach_microscopic_2013,parker_model_2014,berkelbach_microscopic_2014,zeng_low_lying_2014,arago_dynamics_2015,yang_first-principle_2015,tao_understanding_2015,zang_quantum_2017,li_an_2020} Experimentally, the time-resolved spectra techniques, like ultrafast pump-probe,\cite{jundt_exciton_1995,ma_singlet_2012,herz2015unveiling,monahan2017dynamics} two-dimensional electronic spectroscopies,\cite{bakulin_vibrational_2015,breen2017triplet} and two-photon photoemission spectroscopies,\cite{chan_observing_2011,chan_energy_2012,monahan2017dynamics} were gradually employed to study the SF reaction of different systems, including thin films or crystals of tetraene, pentacene, and diphenylisobenzofuran, etc. The relevant measurements and analyses largely identify that the multiexciton state plays a potential role in the SF process. In addition, the SF dynamics in different systems may take place in various time scales, from femtosecond to picosecond domains.\cite{smith_singlet_2010,smith_recent_2013,casanova_theoretical_2018,miyata_triplet_2019,dimitriev_dynamics_2022}

Theoretically, with the development of high-level quantum chemistry theory, the precise electronic structure calculations can be performed in the SF materials, and the effective model Hamiltonian can be constructed and analyzed.\cite{paci_singlet_2006,greyson_maximizing_2010,greyson_singlet_2010,teichen_microscopic_2012,beljonne_charge-transfer_2013,berkelbach_microscopic_2013-1,berkelbach_microscopic_2013,parker_model_2014,berkelbach_microscopic_2014,zeng_low_lying_2014,arago_dynamics_2015,yang_first-principle_2015,tao_understanding_2015,zang_quantum_2017,schroeder_tensor_2019,li_an_2020,zhao_minimal-active-space_2021} For instance, the relative energetics of the low-lying states for the pentacene monomer were once investigated by various electroinc structure methods.\cite{smith_singlet_2010,smith_recent_2013,casanova_theoretical_2018,miyata_triplet_2019,dimitriev_dynamics_2022} Particularly, the pentacene dimer receives considerable research attention, because it provides a prototype to study the SF mechanism. Different strategies were proposed to construct the diabatic Hamiltonian of pentacene dimer. For instance, the diabatic Hamiltonian matrix elements were formulated by Smith and Michl based on a minimum configuration space;\cite{smith_singlet_2010} the electronic diabatic couplings were approximated by Berkelbach et al. in terms of the Fock matrix elements of the HOMO and LUMO of two pentacene molecules;\cite{beljonne_charge-transfer_2013} the possible nonadiabatic coupling between the single and multiexciton states for pentacene dimers was constructed by Feng et al.;\cite{feng_fission_2013} a five-state model Hamiltonian was constructed by Liang and co-workers in terms of the time-dependent density function theory to investigate the J- and H-aggregate behavior;\cite{zang_quantum_2017} an ab inito exciton model was presented by Mart\'{i}nez and co-workers, in which the Frenkel exciton model was further extended, including valence, charge-transfer, and multiexcitonic excited states.\cite{li_an_2020} 
The SF model Hamilotinian was also built from the multistate density functional theory developed by Gao and co-workers.\cite{zhao_minimal-active-space_2021}


Besides, different dynamics methods from the exact full-quantum to semi-classical ones were used to simulate the SF reaction and explore the possibility of various mechanisms proposed.\cite{troisi_dynamics_2006,greyson_maximizing_2010,teichen_microscopic_2012,chan_quantum_2013,berkelbach_microscopic_2013-1,berkelbach_microscopic_2013,zarea_decoherence_2013,berkelbach_microscopic_2014,tao_bath_2014,tao_electronically_2014,akimov_nonadiabatic_2014,tao_understanding_2015,trinh_intra_2015,strong_tetracene_2015,zheng_ultrafast_2016} For example, several groups including Ratner\cite{greyson_maximizing_2010,zarea_decoherence_2013}, Eaves\cite{teichen_microscopic_2012}, Reichman,\cite{chan_quantum_2013,berkelbach_microscopic_2013-1,berkelbach_microscopic_2013,berkelbach_microscopic_2014} and Grozema\cite{mirjani_theoretical_2014,renaud_intermolecular_2015} used the quantum master equation method based on the reduced density matrix to explore the intrinsic mechanism behind the SF reaction. Zhao and co-workers once employed the stochastic Schr\"odinger equation to study the SF dynamics.\cite{zang_quantum_2017}  
The ultrafast nonadiabatic dynamics for the SF process were also extensively investigated by using the quantum dynamics methods that treat all involved electronic and nuclear degrees of freedom explicitly.\cite{tamura_first-principles_2015,zheng_ultrafast_2016,reddy_quantum_2019,schroeder_tensor_2019,mardazad_quantum_2021} Different quantum dynamics approaches were taken, such as the multilayer multiconfigurational time-dependent Hartree method by Burghardt, Tamura and co-workers,\cite{tamura_first-principles_2015} Thoss and coworkers\cite{reddy_quantum_2019} and ourselves\cite{zheng_ultrafast_2016}, and the tensor network by Chin and co-workers\cite{schroeder_tensor_2019}, Ma and co-workers.\cite{mardazad_quantum_2021} The trajectory-based method is also a practical way to treat the SF dynamics in complex system. For instance, Tao used the symmetrical quasi-classical dynamics method to study the influence of different conditions on the SF dynamics;\cite{tao_bath_2014,tao_electronically_2014,tao_understanding_2015} Akimov and co-workers used the trajectory surface hopping method to investigate the nonadiabatic dynamics of charge transfer and SF at the pentacene/C$_{60}$ interface.\cite{akimov_nonadiabatic_2014}

With the help of these important pioneer explorations, the basic knowledge framework about the SF process is constructed.\cite{smith_singlet_2010,smith_recent_2013,casanova_theoretical_2018,miyata_triplet_2019,dimitriev_dynamics_2022} There is no doubt that the intermediate multi-exciton state takes extremely important roles in the SF reaction, thus, its basic properties, dissociation and formation processes are usually the concentration of many studies. The previous works reveal that the involvement of the multi-exciton state is essential to the SF dynamics, which is an entangled triplet-state pair $TT$, i.e. two triplets coupling with each other to form an overall spin singlet. 

For the formation mechanism of the $TT$ state, it has long been the focus of scientific debates.\cite{smith_singlet_2010,smith_recent_2013,casanova_theoretical_2018,miyata_triplet_2019,dimitriev_dynamics_2022} To date, it is widely accepted that the $TT$ formation is highly system-specific, and two mechanisms are commonly used for understanding the intrinsic characteristic involved in the SF reaction, especially for the short-time dynamics. One is the direct mechanism via one two-electron process, in which the local excited state $LE$ is directly decayed into the $TT$ state. The other is the mediated mechanism in terms of two one-electron processes, in which the intermediate charge transfer state $CT$ acts as a bridge for the transformation from the $LE$ state to the $TT$ state, and plays a significant regulatory role in the SF reaction.

Although the existing experimental and theoretical advances provide many evidences for inferring or different perspectives for understanding the inherent mechanism behind the SF process, it is still not enough to uncover overall microscopic physical insights in the SF reaction. As research further continues, it is important to study the rather complex many-body problem involving the interplay between vibrational and electronic degrees of freedom in the SF dynamics. This target requires two efforts. First, it is necessary to build reasonable models, which include all relevant electronic states (the $LE$, $CT$ and $TT$ states) and important nuclear degrees of freedom. Preferably, all parameters in the model should be very accurate to describe realistic systems, which should be derived from the high-level electronic structure calculations. Second, the accurate quantum dynamics study should be performed to treat all nuclear and electronic degrees of freedom with equal root. By doing so, it is possible to provide a very detailed understanding of coupled electronic-nuclear motions in the SF dynamics of realistic systems. 


In this paper, we try to provide a rather accurate description of the SF process in the pentacence dimer by using the electronic-structure and quantum dynamics calculations. First, a linear vibronic coupling  model Hamiltonian was constructed, which includes five electronic states (two $LE$ states, two $CT$ states, one $TT$ state) and all intramolecular vibrational degrees of freedom of the system. Since the diabatization is not an easy task for the current system, we employ the five-state electronic Hamiltonian built from the state averaged extended multi-configurational quasi-degenerate perturbation theory (XMCQDPT) by Zeng, Hoffmann and Ananth.\cite{zeng_low_lying_2014} Next, the additional high-level extended multi-state complete active space second order perturbation theory (XMS-CASPT2)\cite{shiozaki2011communication} was employed to get the vibronic coupling parameters in the diabatic Hamiltonian. Finally, the full quantum dynamics of the SF dynamics was simulated by using the recently-developed tensor train decomposition method.  
In the tensor train format,\cite{schollwock_density-matrix_2011,paeckel_time-evolution_2019} the quantum state and operator are expressed into different matrix products, respectively. Since the high-order tensors are decomposed into a series of localized low-order and low-rank ones, the curse of dimensionality can be greatly alleviated, and the computational costs can also be largely reduced. Therefore, the tensor train approach should be a very efficient and powerful algorithm to simulate the full quantum evolution of complex many-body systems with many coupled degrees of freedom. In this sense, it is possible to employ such the approach to study the SF dynamics of pentacence dimer. Particularly, this approach allows us to provide a very detailed analysis on the roles of the quantum coherence, vibronic couplings and dissipation effects in the SF dynamics. Overall, this work provides a deep and detailed understanding of the SF dynamics in pentacence dimer and relevant systems. 



\section{II. Theoretical Methods and Computational Details}
\subsection{A. Model Hamiltonian}

\subsubsection{A1. Diabatic Hamiltonian}
In order to effectively discuss the SF dynamics of the pentacene dimer here, the molecular Hamiltonian employed in the present work includes five low-lying singlet excited electronic states, namely, two $LE$ states ($S_0S_1$ and $S_1S_0$), two $CT$ states ($CA$ and $AC$), and one triplet-triplet coupled singlet state $TT$. The nuclear motion was described by a set of dimensionless normal modes, and all intramolecular vibrational degrees of freedom were retained. The linear vibronic-coupling Hamiltonian was considered to construct the diabatic model.\cite{domcke2004conical}
In the current case, the total Hamiltonian $\hat{H}_{d}$ is written as the summation of the kinetic operator and diabatic potential Hamiltonian, namely,

\begin{align}
&\hat{H}_{d} = \hat{T} + \hat{V}_{d} \notag \\
&\hat{T}  =  \sum_i^{N_b} \frac{1}{2} \omega_{i} \hat{P}_{i}^2 \notag \\
&\hat{V}_d  =
\sum_k\sum_l\left | \psi_k \right >V_{kl} \left <\psi_l\right |, \\
&V_{kk} = V_{k}^{(0)} + \sum_i^{N_b} \frac{1}{2} \omega_{i} Q_{i}^2 + \sum_i^{N_b} \kappa_{i}^{(k)}Q_{i}, \notag
\end{align}
where $\psi_k(\psi_l)$ denotes the electronic state involved. $V_{kk}$ represents the corresponding energy of $\psi_k$, $V_{kl} (k\neq l)$ is the interstate diabatic coupling between $\psi_k$ and $\psi_l$. $\hat{P}_{i}$ and $Q_{i}$ are the corresponding momentum operator and position of the vibrational mode with frequency $\omega_i$, respectively, under the dimensionless normal coordinates. $V_k^{(0)}$ is the vertical energy of the diabatic state. $\kappa_{i}^{(k)}$ characterizes the vibronic coupling between the $k$-th electronic state and the $i$-th mode. $N_b$ represents the number of vibrational modes.


                                                                              
\subsubsection{A2. Diabatic electronic states and their couplings}

The parameters in the diabatic Hamiltonian, including the energies of different electronic states ($V_{k}^{(0)}$) at $Q_i=0$ and the diabatic couplings ($V_{kl}$), were directly taken from the previous work by Zeng et al.,\cite{zeng_low_lying_2014} and their values can be found in Table 1 of Appendix. The reason is that the electronic Hamiltonian in their work was obtained by the XMCQDPT method with high accuracy, which gives the excellent explanation of experimental results and agrees well with other theoretical simulations. It is worth noting that the energy difference between $CA$ and $AC$ states is rather large, which is due to the asymmetric geometrical alignment of the two monomers. This indicates that $CT$ states may play different roles in the SF dynamics. The ground state $S_0S_0$ is well separated from other electronic states, thus 
this state is not included.


Again we wish to emphasize that the construction of the diabatic model in the current system is not a trivial task, particularly as the model involves five electronic states and the $TT$ state shows the double excitation characters. As the model by Zeng et al. gives the good description of the SF dynamics of the current system, we decide to simply take it. The possible influence of the parameters in the model Hamiltonian on the SF dynamics will be discussed in the below section.

\subsubsection{A3. Vibronic couplings}
On this basis, all parameters for the vibronic couplings were obtained with the assistance of our own high-level electronic structure calculations. 
To simplify the model, we only consider the intramolecular vibrational modes in the construction of the diabatic Hamiltonian. The vibrational modes of each monomer were used as the basis to characterize vibronic couplings. From the monomer, the normal mode analysis at the ground state minimum was performed and all modes were included. 

The vibronic coupling constants $\{\kappa_{i}^{(k)}\}$ are obtained by the transformation of the gradient from Cartesian coordinates to dimensionless normal ones,\cite{domcke2004conical} i.e.

\begin{align}
  \bm{\kappa} = -\boldsymbol{f} \boldsymbol{L},
\end{align}
where $\boldsymbol{f}$ is the corresponding force matrix acting on each atom in Cartesian coordinates on the relevant excited state at the equilibrium geometry of monomer, $\boldsymbol{L}$ is the transformation matrix between Cartesian and dimensionless normal coordinates. Different approaches were used to obtain the vibronic coupling terms for different states.

For the $LE$ state, only one molecular in the pentancene dimer is excited, thus the gradient of the monomer in the excited state just needs to be calculated. For the modes belong to the same monomer with the excitation, the $\kappa_i^{(k)}$ values are constructed by the formula above. For the modes belong to the other momoner lying on the ground state, their $\kappa_i^{(k)}$ are naively set to zero.

For the $CT$ states, both $CA$ and $AC$ can be viewed as a combination of the positive ion ($C$) and negative one ($A$). With this assumption, the gradients of  positive and negative ions were calculated to get $\kappa_i^{(k)}$ for the corresponding ion fragment, respectively. Then, we simply combine them together to give all vibronic coupling terms of the $CT$ states. 

For the $TT$ state, it can also be treated as two triplet states on the respective monomer, and the same strategy can be used to approximate the vibronic coupling.

There is no doubt that the above method is certainly an imperfect one for generating all vibronic coupling terms, since the intermolecular interaction between monomers is missing. The current construction strategy is adopted for two reasons. On the one hand, all intermolecular vibrational modes with the very low frequencies should show rather large anharmonicity, while the current approach is not accurate enough to obtain the precise description of these modes. On the other hand, the SF dynamics of the pentacene dimer takes place very quickly, thus these low-frequency intermolecular vibrational modes should only play a very minor role. Certainly, the influence of these modes can be further included by the introducing the static disorder. Overall, the current  model should already capture all necessary features to treat the ultrafast SF dynamics of the pentacene dimer. 

\subsection{B. Quantum dynamics}                    

\subsubsection{B1. Tensor train method}

In the tensor-train framework,\cite{paeckel_time-evolution_2019} the matrix-product state (MPS) can be used to reconstruct the quantum state:

\begin{align}
  \left | \Psi \right >
  = \sum_{\{s_i\}} \boldsymbol{A}^{s_1}\cdots\boldsymbol{A}^{s_i} \cdots \boldsymbol{A}^{s_m} \left | s_1\cdots s_i\cdots s_m \right >,
\end{align}
where the set $\{\left |s_i \right >\}$ includes $m$ local bases, and the site-dependent tensor $\boldsymbol{A}^{s_i}$ has the $\alpha_{i-1}\times s_i\times \alpha_i$ ($\alpha_0 = \alpha_m = 1$) dimension. Due to the invariance of gauge transformation, the MPS possesses diverse representations, including the left, right, and mixed canonical forms:

\begin{align}
  \left | \Psi \right >
  &= \sum_{\{s_i\}} \boldsymbol{L}^{s_1}\cdots\boldsymbol{L}^{s_i} \cdots \boldsymbol{L}^{s_m} \left | s_1\cdots s_i\cdots s_m \right >, \notag \\
  &= \sum_{\{s_i\}} \boldsymbol{R}^{s_1}\cdots\boldsymbol{R}^{s_i} \cdots \boldsymbol{R}^{s_m} \left | s_1\cdots s_i\cdots s_m \right >, \\ 
  &= \sum_{\alpha_{i-1}, s_i, \alpha_i}[\boldsymbol{M}^{s_i}]_{\alpha_{i-1}, \alpha_i} 
     \left | \Psi_{L, \alpha_{i-1}}^{[1:i-1]} \right >\left | s_i \right > \left | \Psi_{R, \alpha_i}^{[i+1:m]} \right >, \notag
\end{align}
where $\boldsymbol{L}^{s_i}$ and $\boldsymbol{R}^{s_i}$ represent the left and right orthonormal tensors, respectively; $\boldsymbol{M}^{s_i}$ is the $i$th-site center block; $\left | \Psi_{L, \alpha_{i-1}}^{[1:i-1]} \right >$ and $\left | \Psi_{R, \alpha_{i}}^{[i+1:m]} \right >$ corresponds to the intermediate states generated in the left and right orthonormal representations, respectively.

Following an analogous strategy, the matrix product operator (MPO) can also be used to reformulate the relevant operator:

\begin{align}
  \hat{O}
  = \sum_{\{s_i\},\{s_i^{\prime}\}, \{\beta_i\}} \boldsymbol{W}^{s_1, s_1^{\prime}} \cdots \boldsymbol{W}^{s_i, s_i^{\prime}}
    \cdots \boldsymbol{W}^{s_m, s_m^{\prime}} \left | s_1\cdots s_i\cdots s_m \left > \right < s_1^{\prime}
    \cdots s_i^{\prime}\cdots s_m^{\prime} \right |,
\end{align}
where the site-dependent tensor $\boldsymbol{W}^{s_i, s_i^{\prime}}$ has the $\beta_{i-1}\times s_i \times s_i^{\prime} \times \beta_i$ ($\beta_0 = \beta_m = 1$) dimension. Since the local basis can be chosen in different formats, the MPO itself is not always unique, which leads to some different construction strategies developed in recent years. In current work, the electron-phonon Hamiltonian was first reformulated in terms of the second quantization, then the occupation number representation was used to construct the MPO.

By means of the above matrix product representations, several algorithms were proposed to propagate the time-dependent Schr$\rm\ddot{o}$dinger equation (TDSE). In the current work, the time-dependent variational principle (TDVP) method was employed to simulate the dynamics process and calculate the corresponding spectral signals, due to its very high efficiency and accuracy.\cite{paeckel_time-evolution_2019} Overall, the TDVP algorithm realizes that the evolution of the quantum state is conducted in a confined MPS manifold, and the specific mathematical details can be referred to the important articles and reviews. While in implementation, the original TDSE was transformed and solved by following two equations:

\begin{align}
  i\hbar\frac{d}{dt} \left | \Psi(t) \right >
  &= \sum_{i=1}^m \hat{P}_{i-1}^L \otimes \hat{I}_i \otimes \hat{P}_{i+1}^R\hat{H} \left | \Psi(t) \right >, \notag \\
  i\hbar\frac{d}{dt} \left | \Psi(t) \right >
  &= -\sum_{i=1}^{m-1}\hat{P}_{i}^L \otimes \hat{P}_{i+1}^R\hat{H} \left | \Psi(t) \right >,
\end{align}
where $\hat{P}_i^L$ and $\hat{P}_i^R$ are the left and right projectors, respectively, and satisfy:

\begin{align}
  &\hat{P}_i^L 
  =  \sum_{\alpha_i} \left | \Psi_{L, \alpha_{i}}^{[1:i]} \right > \left < \Psi_{L, \alpha_{i}}^{[1:i]} \right |, \\
  &\hat{P}_i^R 
  =  \sum_{\alpha_i} \left | \Psi_{R, \alpha_{i-1}}^{[i:m]} \right > \left < \Psi_{R, \alpha_{i-1}}^{[i:m]} \right |. \notag
\end{align}
Thus, one needs to individually and sequentially calculate $m$ forward-propagation and $m-1$ backward-propagation equations in a single step propagation.

\subsubsection{B2. Initial conditions}
\noindent
In the time-dependent dynamics propagation, the system was assumed to be prepared by vertical placing the lowest vibrational level of the ground state to one of local excited states or their superposition, i.e.

\begin{align}
    \left . \left |\Psi(t = 0) \right > \right > = \left [ c_{S_0S_1}\left |\Psi_{S_0S_1} \right >+c_{S_1S_0}\left |\Psi_{S_1S_0} \right > \right ] \left | \textbf{0} \right >,
\end{align}
where $c_{S_0S_1}$ and $c_{S_1S_0}$ are the coefficients corresponding to the electronic wave functions $\left |\Psi_{S_0S_1} \right >$ and $\left |\Psi_{S_1S_0} \right >$, respectively. The lowest vibrational state $\left |\textbf{0} \right >$  on the ground electronic state is given as 

\begin{align}
    \left | \textbf{0} \right > = \left |0, \cdots, 0 \right >.
\end{align}

Finally, the electronic population and coherence can be obtained by the construction of the reduced density matrix for electronic motions, which is calculated by the trace of the total density matrix over all nuclear degrees of freedom, i.e.

\begin{align}
  \rho_{ij}(t) = Tr_n\{\left | \psi_i \right > \left < \psi_j \large | \Psi(t) \right >\left < \Psi(t) \right | \}
\end{align}

\subsection{C. Computational details}
\noindent
In the current work, the reference geometry of pentacene dimer was directly taken from the previous work, and the corresponding structures are shown in Fig. 1. In fact, different stacked geometries were also found in the crystal. However, the SF dynamics of different stacked dimers normally show a similar time scale of the ultrafast SF dynamics, although their dynamics details are different. Thus, it is enough to take a represent one to conduct the detailed study on the SF dynamics. The results should give a deep understanding of the mechanism of the SF dynamics of the current stacked pentacene dimer, as well as providing the important hints to explore the SF processes in more general stacked configurations.

\begin{center}
  \begin{figure}[htbp]
      \centering
      \includegraphics[width=6cm,height=6cm]{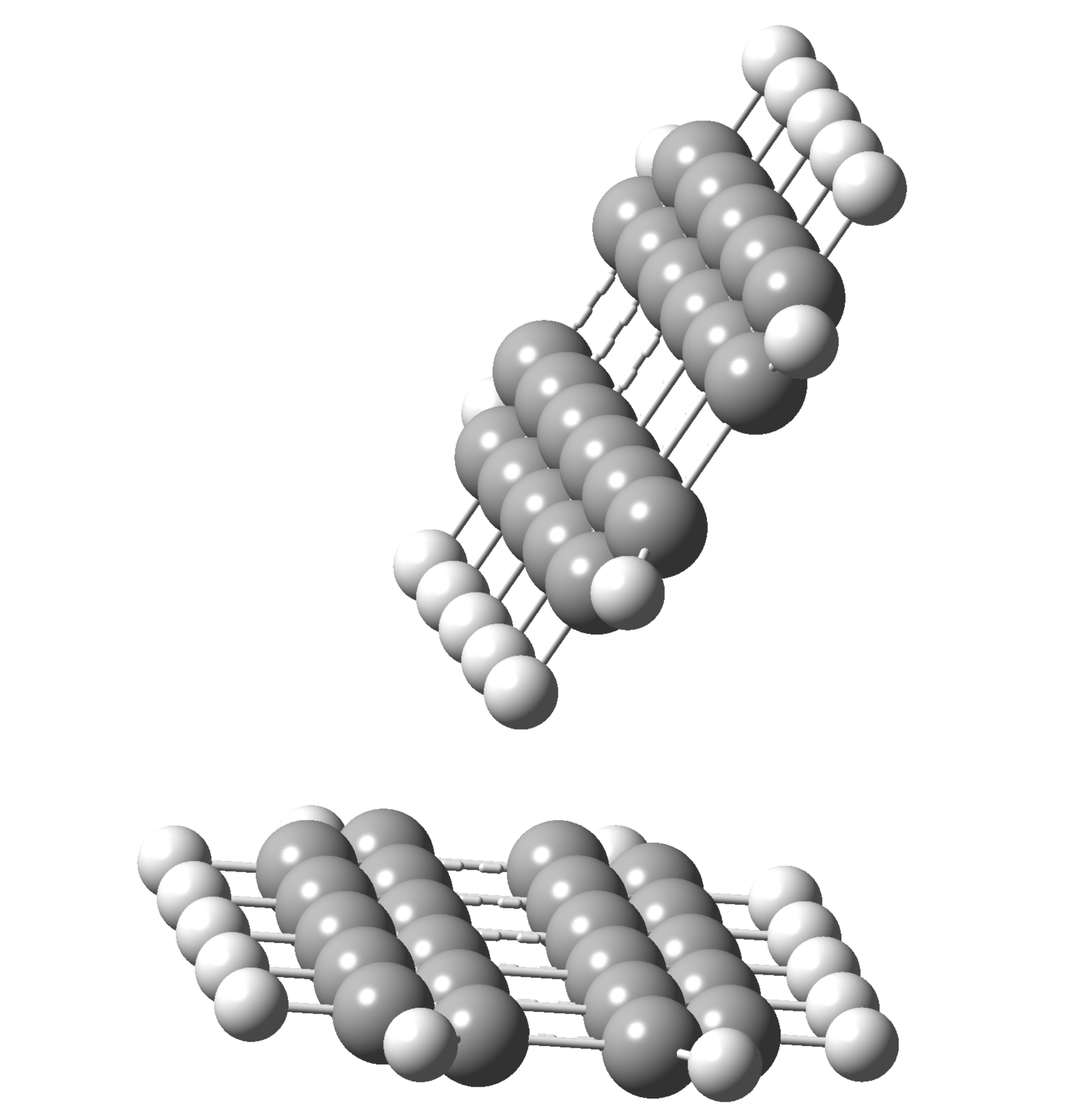}
      \caption{The structure of pentacene dimer.}
  \end{figure}
\end{center}

In order to obtain the vibronic-coupling parameters, the geometry optimization and vibrational analysis of the ground-state minimum for the pentacene monomer were performed at the MP2/CCPV-DZ level using Gaussian 16 software.\cite{frisch2016gaussian} This step gave us the frequency and dimensionless normal coordinates of each mode at the ground-state minimum. Then, the gradients of cation, anion, singlet excited and triplet states at the equilibrium structure of pentacene monomer were calculated with the XMS-CASPT2/CCPV-DZ level using BAGEL program,\cite{shiozaki2018bagel} in which the active space of 12 electrons in 12 orbitals were employed, all active orbitals are $\pi$ and $\pi^*$ orbitals, and three states were averaged to obtain the excited state information.

In the SF dynamics propagation, the initial wave packet at $t = 0$ was directly placed to two $LE$ states 
($\left | \Psi_{S_0S_1} \right >$ or $\left | \Psi_{S_1S_0} \right >$), as well as their symmetric or asymmetric combinations 
($\frac{1}{\sqrt{2}}\left [ \left |\Psi_{S_1S_0}\right >+\left |\Psi_{S_0S_1}\right > \right ]$ or $-\frac{1}{\sqrt{2}}\left [ \left |\Psi_{S_1S_0}\right >-\left |\Psi_{S_0S_1}\right > \right ]$).
The SF quantum dynamics was simulated by the tensor-train method. To construct the MPS and MPO effectively, the truncated value was set to $10^{-13}$ for single value decomposition, and the number of basis set for vibrational modes was obtained by the convergence test. Finally, the whole dynamics was evolved to 400 fs with the time step of 0.5 fs.


\section{III. Results and Discussion}
\subsection{A. Normal modes and vibronic couplings}
For the pentacene monomer, totally  102 normal modes were  obtained from the frequency analysis at the ground-state minimum.


In order to better characterize the importance of different normal modes, Fig. 2 shows the $\kappa / \omega$ values of normal modes for the singlet excited, cationic, anionic, and triplet states of the pentacene monomer. Three groups of molecular vibrations with different frequency domains, < 0.05 eV, 0.05-0.25 eV, > 0.25 eV, were examined. 

As  seen, most normal modes with the frequencies lower than 0.05 eV display very small $|\kappa| / \omega$ values, which means that the effects of low-frequency normal modes may not be important. Only two low-frequency modes show large vibronic couplings for some electronic states. Fig. 3(a-b) show the vibrational motions of these two normal modes in detail, and the main movement here involves the in-plane stretching and bending motions of the pentacene skeleton.

When the frequency reaches 0.05-0.25 eV, several normal modes exhibit strong vibronic couplings, and thus they should have the significant impact on the nonadiabatic SF dynamics. And the vibrations of two typical normal modes within this frequency range are collected in the Fig. 3(c-d), in which the change of the skeleton is also the leading way of movement. Particularly, most modes with strong vibronic coupling are relevant to the CC bond stretching motions. The reason is easily to be understood. Here all involved states display the $\pi\pi^*$ characters. The CC bond stretching motion can modify the conjugated statues, changing the energies of $\pi$ and $\pi^*$ orbitals. As the consequence, the $\pi\pi^*$ state energy is highly dependent on the CC bond lengths. Therefore, these modes display the strong vibronic couplings.     

The remaining high-frequency normal modes with frequency larger than 0.25 eV exhibit small $|\kappa| / \omega$ values, and the main motion is the in-plane stretching motion of the carbon-hydrogen bond, as shown in Fig. 3(e). Since the CH distances have the little influence on the conjugated statues and therefore do not modify the  
energies of the $\pi\pi^*$ states. Therefore, these modes display small vibronic couplings

\begin{center}
  \begin{figure}[htbp]
      \centering
      \includegraphics[width=16cm,height=14cm]{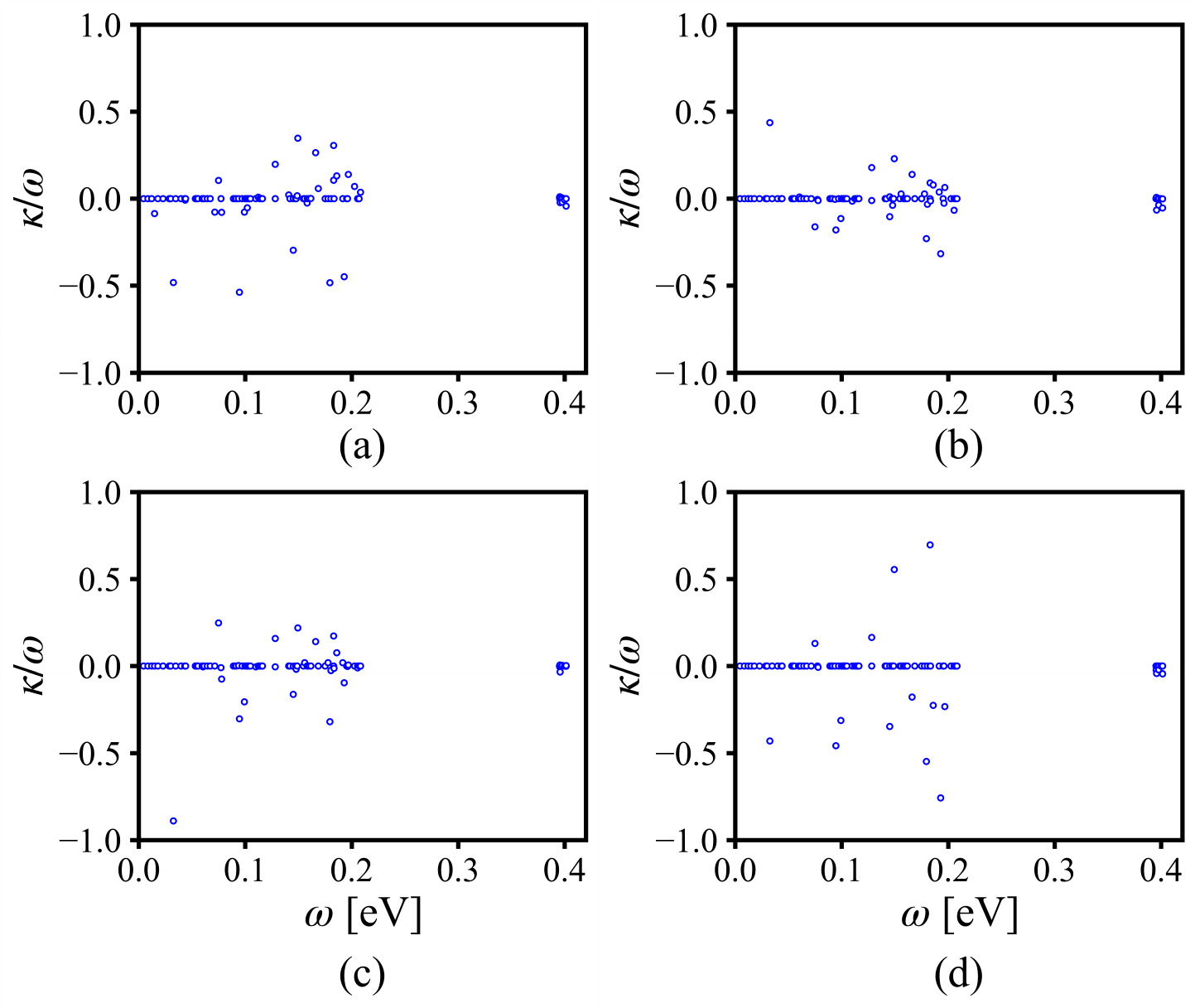}
      \caption{The $\kappa / \omega$ values of different dimensionless normal mdodes for the pentacene monomer: (a) singlet excited state, (b) cationic state, (c) anionic state, and (d) triplet state.}
  \end{figure}
\end{center}

\begin{center}
  \begin{figure}[htbp]
      \centering
      \includegraphics[width=16cm,height=5.7cm]{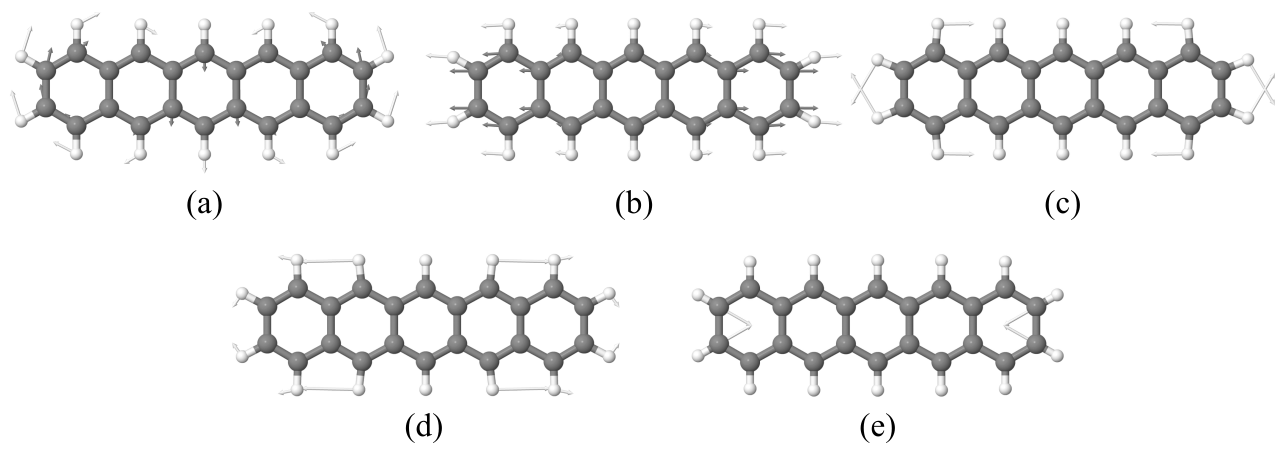}
      \caption{The vibration directions of some typical normal modes for the pentacene monomer: (a) corresponding to the frequency with 0.014729 eV (b) corresponding to the frequency with 0.032516 eV, (c) corresponding to the frequency with 0.144957 eV, (d) corresponding to the frequency with 0.149243 eV, (e) corresponding to the frequency with 0.401401 eV.}
  \end{figure}
\end{center}

\subsection{B. Pure electronic dynamics}
\noindent
Then, we consider the pure electronic dynamics without the vibrational degrees of freedom. Although such dynamics in principle are rather simple, we still prefer to start our discussion from them, because the careful analyses on these data may display some deep physical insights to help up for the further understanding of the true nonadiabatic SF dynamics. 

The pure electronic dynamics starting from different initial conditions are shown in Fig. 4, in which the obvious population transfer occurs among different diabatic electronic states. As expected, the oscillation appears in the population dynamics, which reflects the quantum coherence in the pure electronic dynamics. In fact, starting from different initial conditions, the population transfer dynamics here show different patterns. When the initial state is $-\frac{1}{\sqrt{2}}\left [ \left |\Psi_{S_1S_0}\right >-\left |\Psi_{S_0S_1}\right  > \right ]$, the small amplitude population oscillation is observed, and the oscillation period is rather long with respect to other cases. When the initial state is $\left |\Psi_{S_0S_1}\right >$ or $\left |\Psi_{S_1S_0}\right >$, the oscillation in the population becomes significant, indicating the fast population transfer between two $LE$ states. When the initial state $-\frac{1}{\sqrt{2}}\left [ \left |\Psi_{S_1S_0}\right >+\left |\Psi_{S_0S_1}\right  > \right ]$ is prepared, we noticed that the high-frequency oscillations appear in the population dynamics and many states are populated. These results indicate that the full understanding of the pure electronic dynamics is not trivial. For the further analysis, we performed the fast Fourier transform (FFT) on the time-dependent population, and the diabatic-to-adiabatic transformation matrix of the electronic Hamiltonian was checked to explain some interesting features. The concrete results are summarized in Fig. 5 and 6.

For the initial state with $-\frac{1}{\sqrt{2}}\left [ \left |\Psi_{S_1S_0}\right >-\left |\Psi_{S_0S_1}\right > \right ]$, the dynamics is basically characterized by the coincident population oscillation between two $LE$ states, while other states seem not to be populated. The overall oscillation patterns look very simple in the population dynamics. 
The FFT analyses show that the position of the highest peak approximates 66.67 fs, which corresponds to the oscillation frequency with the energy difference between adiabatic electronic states $S_2^a$ and $S_3^a$. This phenomenon can also be understand in terms of the diabatic-to-adiabatic transformation matrix shown in Fig. 6, where both $S_2^a$ and $S_3^a$ states have larger weights.
Although in the first glance, the quantum dynamics here seems to only involve two $LE$ states, and other states are not relevant at all. However, this is not true since such an oscillation can not be the simple Rabi-type oscillation between two $LE$ states due to their extremely weak coupling (-0.006 eV). Instead, this population transfer process is fully governed by the super-exchange mechanism mediated by the $CT$ state, in which the $AC$ state is not far from two $LE$ states and also displays the visible $AC$-$LE$ couplings.  This idea is supported by the existence of another two relatively significant peaks at 10.00 and 30.00 fs in the FFT analyse of the population dynamics, which are resulted from the $S_1^a$-$S_3^a$ and $S_3^a$-$S_4^a$ energy differences, respectively. Here, the $S_4^a$ state is mainly characterized by the $AC$ component according to the diabatic-to-adiabatic transformation matrix.

However, for other three cases with $\frac{1}{\sqrt{2}}\left [ \left |\Psi_{S_1S_0}\right >+\left |\Psi_{S_0S_1}\right > \right ]$, $\left | \Psi_{S_0S_1} \right >$ or $\left | \Psi_{S_1S_0} \right >$ as the initial state, the more significant population transfer and large oscillations appear in the population dynamics. The high-frequency components appear in all these cases. The $AC$ state and even the $TT$ state are populated during the electronic dynamics. After the FFT analysis of the electronic populations v.s. time, three peaks are found at the same locations, which are similar to those in the dynamics from the first initial condition with $-\frac{1}{\sqrt{2}}\left [ \left |\Psi_{S_1S_0}\right >-\left |\Psi_{S_0S_1}\right > \right ]$. In addition, several additional peaks were discovered, near 3.85, 9.00, 50.00 fs, which correspond to the $S_3^a$-$S_5^a$, $S_1^a$-$S_4^a$ and $S_1^a$-$S_2^a$ energy differences, respectively. This indicates that two $LE$ components, $AC$ and $TT$ components are involved, while the $CA$ component does not play essential roles. 

According to the electronic Hamiltonian, the $LE$-$LE$ coupling is rather weak and and $LE$-$TT$ coupling is even negligible, while the $LE$-$AC$ and $AC$-$TT$ couplings are visible. Therefore, we believed that the $AC$ state is very important here, which not only bridges for the $LE$-$LE$ states population transfer, but also the $LE$-$TT$ population transfer. Since the energy of the $AC$ state is higher than the $LE$ and $TT$ states, the excited-state energy transfer and the SF processes are governed by the super-exchanging mechanism in the pure electronic dynamics.  The $TT$ state can be formed via the $AC$ state due to the pronounced $AC$-$TT$ coupling, instead of the direct pathway.
The $CA$ state seems not important because of its higher energy. This again shows that the asymmetrical configuration of the stacked dimer bring the potential influcence on the final SF dynamics, because such feature causes different energies of two $CT$ states. 

One central question is why different population dynamics exist for different initial conditions, particular for the symmetric or asymmetric combination of two $LE$ states. The underlying reason is as follows. Two $LE$-$AC$ couplings display the same sign. When we considered the symmetric and asymmetric combinations of two $LE$ states, the resulting state should show the large and small electronic couplings with the $AC$ state, respectively. As the consequence, we should only see the significant and weak population transfer between the symmetric and asymmetric combinations of two $LE$ states and $AC$ one, respectively. This explains different oscillation patterns exist in the population dynamics with different initial conditions. In addition, for the $\frac{1}{\sqrt{2}}\left [ \left |\Psi_{S_1S_0}\right >+\left |\Psi_{S_0S_1}\right > \right ]$ situation, the $TT$ state can be easily formed due to the deep involvement of the $CT$ state in the quantum dynamics. When the dynamics starts from one of the $LE$ state, the population transfer takes place between two $LE$ state due to the $CT$-mediated exchanging mechanism, as discussed above. As the $AC$ state is involved, we also observed the minor formation of the $TT$ state. 
 
\begin{center}
  \begin{figure}[H]
      \centering
      \includegraphics[width=16cm,height=13.56cm]{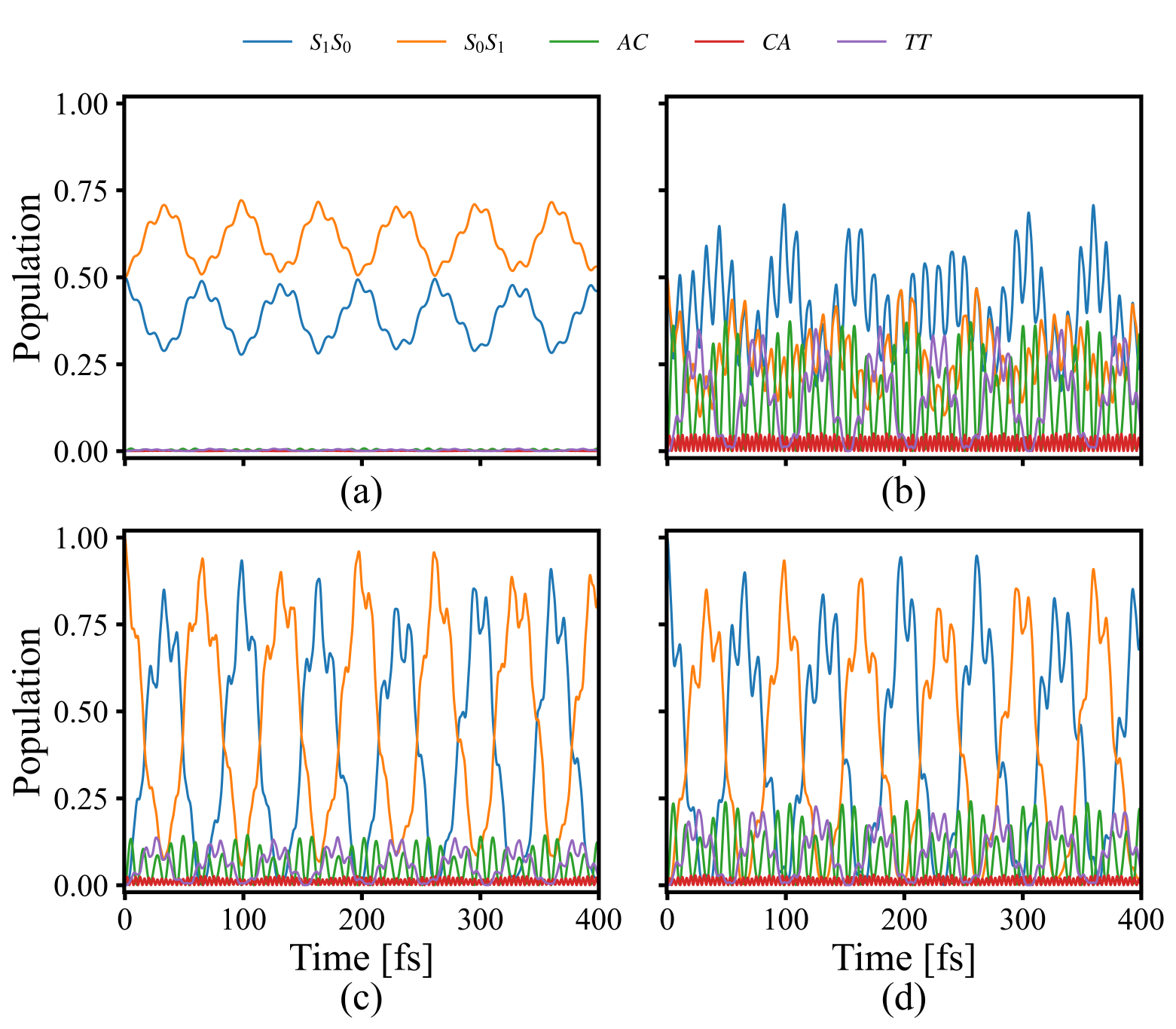}
      \caption{the time-dependent pure electronic population for different initial states: (a) $-\frac{1}{\sqrt{2}}\left [ \left |\Psi_{S_1S_0}\right >-\left |\Psi_{S_0S_1}\right > \right ]$, (b) $\frac{1}{\sqrt{2}} \left [ \left |\Psi_{S_1S_0}\right >+\left |\Psi_{S_0S_1}\right > \right ]$, (c) $\left |\Psi_{S_0S_1}\right >$ and (d) $\left |\Psi_{S_1S_0}\right >$.}
  \end{figure}
\end{center}

\begin{center}
  \begin{figure}[H]
      \centering
      \includegraphics[width=16cm,height=16cm]{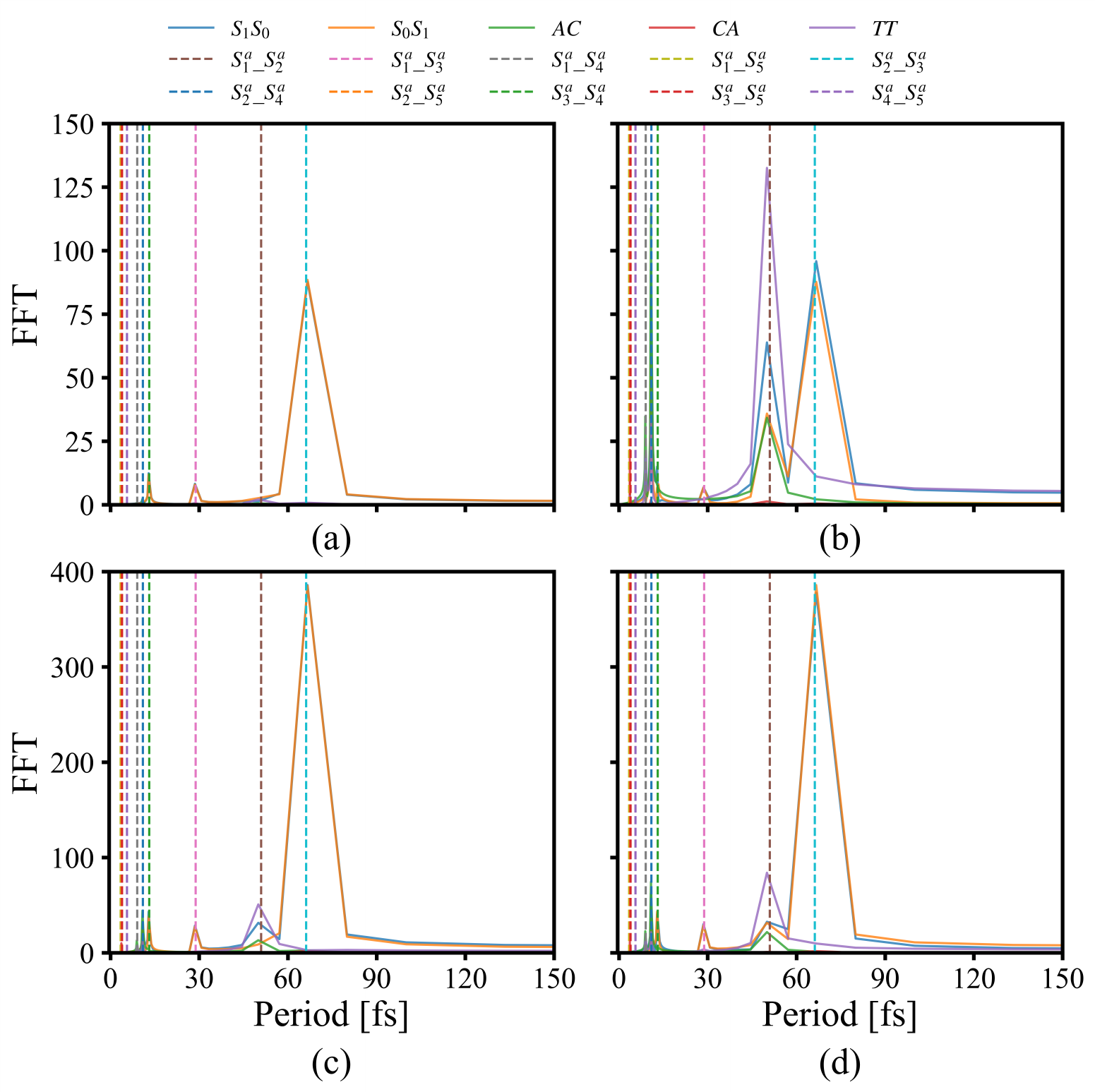}
      \caption{the FFT of electronic population for electronic states $S_1S_0$, $S_0S_1$, electronic $AC$, $CA$ and $TT$ for different initial states: (a) $-\frac{1}{\sqrt{2}}\left [ \left |\Psi_{S_1S_0}\right >-\left |\Psi_{S_0S_1}\right > \right ]$, (b) $\frac{1}{\sqrt{2}} \left [ \left |\Psi_{S_1S_0}\right >+\left |\Psi_{S_0S_1}\right > \right ]$, (c) $\left |\Psi_{S_0S_1}\right >$ and (d) $\left |\Psi_{S_1S_0}\right >$. The vertical dashed lines represent the vibrational frequency corresponding to the adiabatic energy difference between every two states.}
  \end{figure}
\end{center}

\begin{center}
  \begin{figure}[H]
      \centering
      \includegraphics[width=16cm,height=12cm]{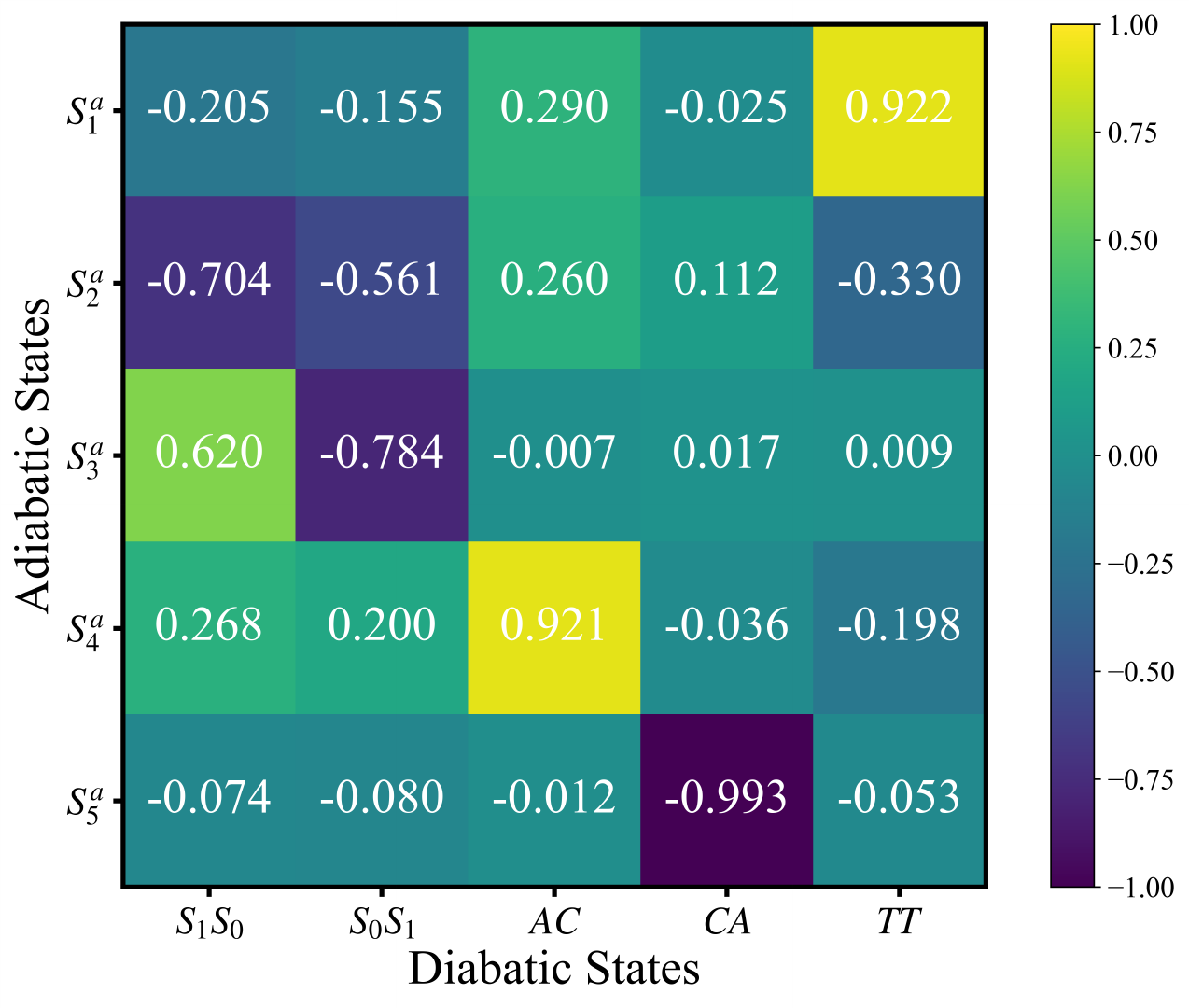}
      \caption{the heatmap of the transformation matrix between diabatic and adiabatic electronic states.}
  \end{figure}
\end{center}

\subsection{C. Singlet fission dynamics}
As the pentacene monomer displays the $C_{2V}$ symmetry, in principle only the $A_1$ modes of each monomer should display the visible linear vibronic coupling terms due to the symmetry selection rule. 
Therefore, the vibrational modes with strong intrastate vibronic coupling here should be the $A_1$ modes that can modify the energy of the $\pi$-$\pi^*$ transitions. In the projection of the gradient, we also find the modes with other symmetries display very minor coupling strengths. Therefore, in the below discussions, we start from the diabatic model with all $A_1$ modes (18 modes for each monomer and totally 36 modes for the dimer). 
Fig. 7 shows the SF dynamics with the inclusion of all effectively vibrational degrees of freedom  from different initial states, in which the time-dependent electronic population exhibits a qualitatively consistent decay pattern. Initially, the population of the $LE$ state decays rapidly in all situations. Simultaneously, the $CT$ state is instantaneously activated, then the  population transfer occurs between the $LE$ and $CT$ states, especially for the initial states with $\frac{1}{\sqrt{2}} [\left |\Psi_{S_1S_0}\right >+\left |\Psi_{S_0S_1}\right > ]$, $\left |\Psi_{S_0S_1}\right >$ and $\left |\Psi_{S_1S_0}\right >$. While throughout the whole dynamics process, the $TT$-state population is increased almost exponentially with the fitted rate constants 120, 70, 106 and 77 fs, respectively, which are close to the previous experimental and theoretical observations.\cite{zang_quantum_2017} Since the $LE$-$TT$ coupling is extremely weak in the current model, the ultrafast SF process is only possible via the $CT$-mediated mechanism.

Similar to the pure electronic dynamics, both $AC$ and $CA$ states take different roles in the whole SF process. Because the $CA$ state lies in a very high energy domain, it is not populated in the SF dynamics.
As the contrast, the population of the $AC$ state is visible, especially when the initial state staring from two $LE$ states or their symmetric combination. Besides, the $AC$ state also displays the obvious population transfer with high oscillation patterns in the early stage of dynamics as shown in Fig. 7.
As the symmetric combination of two $LE$ states may define the strong $LE$-$CT$ couplings, the larger $CT$ population exists for the symmetric combination situations. As a consequence, the formation of the $TT$ state is also slightly faster in this case. The reversed situation is given in the asymmetric combination of two $LE$ states.


In order to further understand the observations above, two reduced four-state models were constructed based on the original model Hamiltonian with the initial state $\frac{1}{\sqrt{2}} [\left |\Psi_{S_1S_0}\right >+\left |\Psi_{S_0S_1}\right > ]$, so that the effect of the $CT$ state can be analyzed in detail. Here, we either discarded the $CA$ or $AC$ state to define two four-state models. 
Fig. A1 in Appendix shows the corresponding time-dependent populations of these reduced four-state models. It can be seen that when only the $AC$ state was considered, the overall dynamics pattern obtained here is quite similar to the original five-state results, although the time-dependent population of electronic state $TT$ is slightly overestimated. While for the second four-state model with the $CA$ state included, the dynamics results are different, and the population transfer to the $TT$ state becomes much slower. Thus, the conclusion can be further supported that the $AC$ and $CA$ states may play different roles for the population transfer of electronic state $TT$. 
Overall, the SF dynamics is prone to follow the population transfer from the $LE$ state to multi-exciton $TT$ one, which is the typical characteristic of the super-exchange mechanism. 

Besides, it is also important to note that the oscillation patterns of the population evolution in the early-stage of the SF dynamics is comparable to the pure electronic dynamics without the consideration of vibrational degrees of freedom, which demonstrates the dominant process at this stage involves the population interchange between the $S_1S_0$ and $S_0S_1$ states mediated by the $CT$ one, and the vibrational degrees of freedom here have little impact on the early-stage dynamics. While with time evolution, the population transfer from the $S_1S_0$ to $S_0S_1$ state appears to be suppressed by vibrational degrees of freedom involved, due to the strong vibronic coupling effects. From this point of view, a natural question would be raised about how  different vibrational degrees of freedom influence the electronic dynamics. This problem was noticed by several groups in their previous studies on the relevant topic.\cite{zheng_ultrafast_2016}

\begin{center}
  \begin{figure}[H] 
      \centering
      \includegraphics[width=16cm,height=14cm]{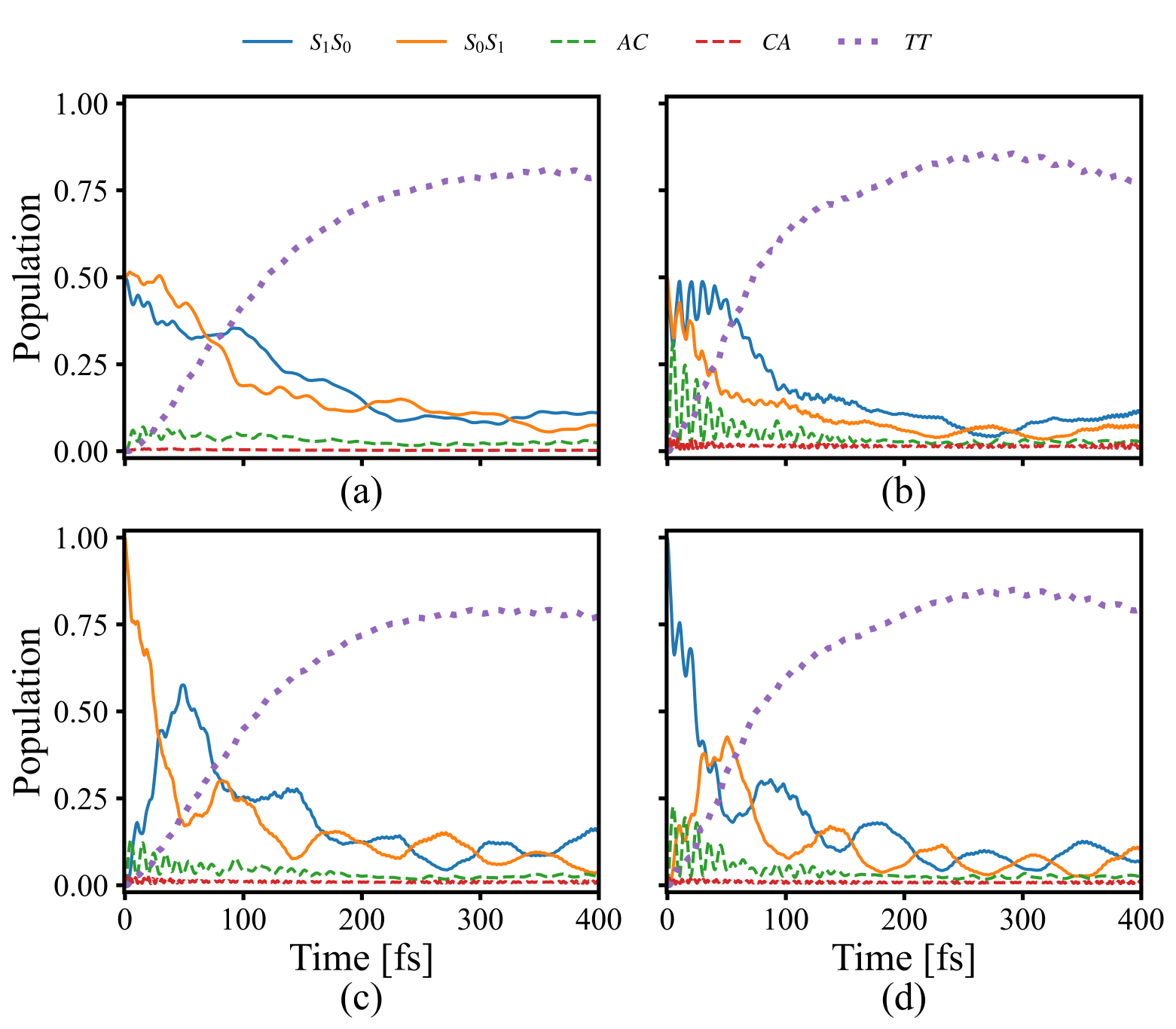}
      \caption{the time-dependent electronic populations of the electronic states $S_1S_0$, $S_0S_1$, $AC$, $CA$ and $TT$ with different initial conditions: (a) $-\frac{1}{\sqrt{2}} [\left |\Psi_{S_1S_0}\right >-\left |\Psi_{S_0S_1}\right > ]$, (b) $\frac{1}{\sqrt{2}} [\left |\Psi_{S_1S_0}\right >+\left |\Psi_{S_0S_1}\right > ]$, (c) $\left |\Psi_{S_0S_1}\right >$ and (d) $\left |\Psi_{S_1S_0}\right >$.}
  \end{figure}
\end{center}

In order to further understand the role of vibrational degree of freedom in the SF dynamics, a series of reduced dimensional Hamiltonian were considered on the basis of the original model.
Then, the contribution of different vibrational modes was investigated in terms of these reduced dimensional Hamiltonian.

In light of the similarities among the SF dynamics patterns above, only the case with the initial state $\frac{1}{\sqrt{2}} \left [ \left |\Psi_{S_1S_0}\right >+\left |\Psi_{S_0S_1}\right > \right ]$ was selected as the prototype for further studies, in order to simplify the problem itself and avoid excessive computations. 

Fig. 8 exhibits the vibrational period of each selected mode, according to this particular distribution, the chosen modes were further divided into eleven parts ($R_1$ -- $R_{11}$) in the following analysis. Therein, the region $R_1$ contains the modes with the vibrational frequencies less than 20 fs, the region $R_{2}$ includes the modes with the vibrational frequencies between 20 and 24 fs, and the remaining modes with the vibrational frequencies larger than 24 fs are sequentially named as the regions $R_{3}-R_{11}$ in accordance with the period from small to large, where each region only contains one vibrational mode for each monomer.
Then, the vibrational modes in different regions were successively considered or combined to perform dynamics calculations, the corresponding population results of the $TT$ state over time are summarized in Fig. 9. 

When only the region $R_{1}$ (total 8 modes with the frequency in 0.397594-0.401401 eV) is considered, where the primary motions here involve the stretching motions  of the carbon-hydrogen bond as shown in Fig. 3(e), the simulated population is quite similar to the pure electronic observation, and the $TT$ population oscillates v.s. time. This indicates that the high-frequency modes or the relative vibrations of carbon-hydrogen bonds have a minor effect on the pure electronic transition, especially in the early stage of dynamics. The underlying reason may consist of two facts, on the one hand, these high-frequency modes are too far from the resonance of the electronic transition (especially from the purple dashed line shown in middle part of Fig. 8); on the other hand, their vibronic couplings are very small as shown in Fig. A2 of Appendix. 

However, when the region $R_{2}$ (total 10 modes with the frequency in 0.179442-0.196718 eV) was used to perform dynamics simulation, a completely different feature is exhibited in Fig. 9. The dynamics here indicates that the vibrational modes in the region $R_2$ can damp the electronic oscillation, thus, the population of electronic state $TT$ becomes flat and stable. This may be due to the fact that the vibronic coupling in this region is much larger and widely distributed as shown in Fig. A2 of Appendix, and the dominant vibration modes here are the change of the pentacene skeleton. 

On this basis, when combined two regions $R_{1}$ and $R_{2}$, only the dynamics pattern appeared in the region $R_{2}$ was left. And an analogous time-dependent population of electronic state $TT$ was obtained when the region $R_{3}$ (total 2 modes with the frequency 0.166037 eV) was included. This indicates that the modes in $R_{3}$  region does not show other effects as compared to the ones in the $R_{2}$ region.

Interestingly, when the region $R_{4}$ (total 2 modes with the frequency 0.149243 eV)
was added into the above combination, the dynamics pattern changes significantly, in which the time-dependent population of the $TT$ state begins to approach the original dynamics results, although the population is underestimated. And this deviation can be greatly alleviated with the involvement of modes in region $R_{5}$ (total 2 modes with the frequency 0.144957 eV), and the time-dependent population of the $TT$ state is very close to the results of the high-dimensional model Hamiltonian.

Thus, the modes in regions of $R_{4}$ and $R_{5}$ have great influences on dynamics. The detailed reasons can be easily understood in the adiabatic picture.  The modes in $R_{4}$ and $R_{5}$ regions are resonant to the oscillation period corresponding to the energy difference between adiabatic states $S_1^a$ and $S_3^a$, whose leading movements are the change of the skeleton as shown in Fig. 3(c-d). 


As for other regions ($R_{6}$--$R_{11}$), the relevant modes involved only play a role in fine-tuning of the dynamics results as shown in Fig. 9. Their minor influcence can be explained by their weak vibronic couplings, and their low frequencies  far from ones for the electronic transitions. 


\begin{center}
  \begin{figure}[H] 
      \centering
      \includegraphics[width=16cm,height=12cm]{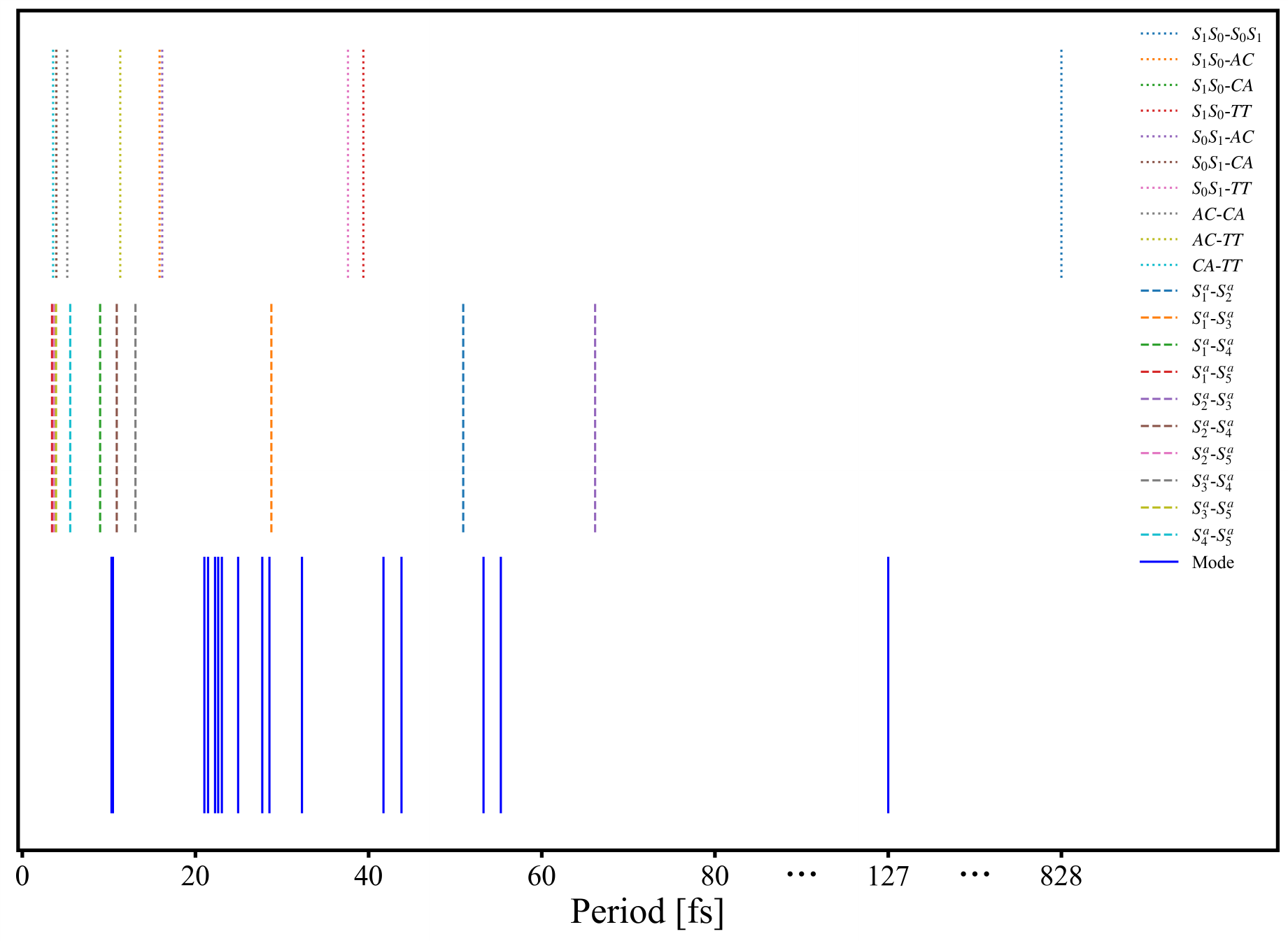}
      \caption{the vibrational period involved in the reduced model, the blue lines denote the vibrational periods of reserved modes, the dashed lines denote the oscillation period corresponding to energy differences between adiabatic states, the dotted lines denote the oscillation period corresponding to energy difference between diabatic states.}
  \end{figure}
\end{center}

\begin{center}
  \begin{figure}[H] 
      \centering
      \includegraphics[width=16cm,height=16cm]{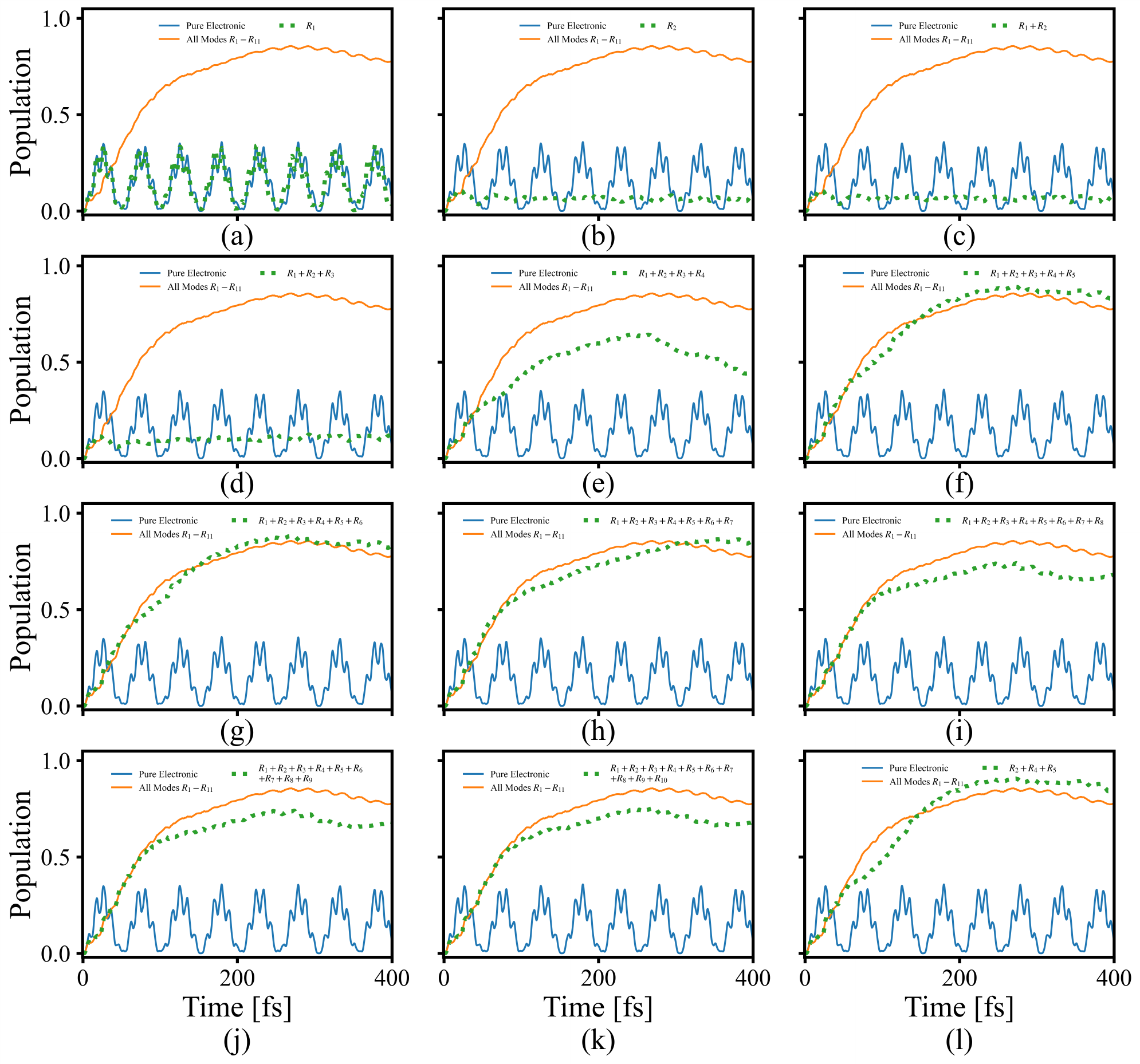}
      \caption{the time-dependent population of electronic state $TT$ based on the initial state $\frac{1}{\sqrt{2}} [\left |\Psi_{S_1S_0}\right > +\left |\Psi_{S_0S_1}\right > ]$ with modes in different regions, in which the modes were combined sequentially in accordance with the period from small to large.}
  \end{figure}
\end{center}

\subsection{D. Discussion}
\noindent
In current work, the model constructed therein combines the data from two different high-level electronic structure calculations. Such a choice is mainly due to the complexity in the construction of the diabatic model. However, we still believe that the current setup is reasonable. On the one hand, the electronic Hamiltonian matrix should be accurate enough since it is constructed at the XMCQDPT level. On the other hand, the electronic states included in the current systems are mainly $\pi\pi^*$ states, and the electronic transitions involve frontier orbitals of $\pi$ and $\pi^*$. In this case, only if some vibrational modes can modify the energies of $\pi$ and $\pi^*$ orbitals, and their interstate vibronic couplings should be strong. In principle, such types of modes should be relevant to the stretching motion of the carbon-carbon bond in the current system. Therefore, even if the vibronic coupling values may depend on the electronic methods used, their major features should be similar, i.e. the important modes and the amplitudes of their vibronic couplings should not far from each other. In this sense, even these numbers changes slightly, we still can obtain the dynamics features, which are not far from the current results. Thus, our current model is reasonable to conduct the studies on the SF dynamics.

One weak point is that the vibrational modes of the monomer were employed to built the vibronic coupling modes. This may bring some bias. For instance, the monomer show the C$_{2V}$ symmetry and therefore only the $A_1$ modes should show the intrastate linear vibronic coupling here. In stacked systems, such symmetry does not strictly remain and therefore the symmetry selection rule should not be followed. However, we expect that the geometrical distortion should be small. In addition, the modes can modify the $\pi$-$\pi^*$ transitions should still be very important here.
As the consequence, we expect that the intramolecular modes that show the strong vibronic coupling should not be similar in the monomer and stacked systems. However, in the stacked systems, the intermolecular vibrational modes should play some roles. This may be treated as the static disorder in the future work.

These is no doubt that the SF process is very complex, thus, more efforts are needed to further explore the dynamics details. Recently, we also notice that some prospecting efforts have been taken into the construction of the SF Hamiltonian, in which the coupling between $LE$ and $TT$ states was further considered, and the SF process was considered in terms of the conical intersection between the singlet exciton and triplet-pair state.\cite{li_an_2020} Thus, it would be very interesting to add the conical intersection into the model directly and check the dynamics performance. 

\section{IV. Conclusion}
In this paper, we make our efforts on the rather accurate and detailed investigation of the SF process in the stacked polyacene dimer by combining the electronic-structure calculations and quantum dynamics simulations. On the basis of the electronic Hamiltonian derived from the previous high-level quantum chemistry method, the vibronic coupling parameters were calculated by ourselves to construct the linear vibronic coupling model. Then, the tensor train decomposition method with high efficiency and great potential was used to study the quantum propagation of the SF process, and explore the influence of different factors.

The model Hamiltonian in this work contains five electronic states (two $LE$, two $CT$, one $TT$) and a bunch of vibrational modes. The frequency analysis shows that several in-plane CC stretching vibrational modes exhibit strong vibronic couplings because they can significantly modify the $\pi$-$\pi^*$ transitions, which have significant impacts on the nonadiabatic SF dynamics.

The pure electronic dynamics  provides some preliminary but important perspectives to understand the SF process. The results indicate that the population transfer dynamics display dissimilar patterns when starting from different initial conditions. The frequency-domain analyses of the population dynamics reveal that the periods of the oscillations of electronic populations are relevant to the energy gap between different adiabatic states. The population transfer between two $LE$ states is fully governed by the super-exchange mechanism mediated by the $CT$ state, and even the $TT$ state can be populated during the electronic dynamics. The symmetric or asymmetric linear combination of two $LE$ states results in different population dynamics because of the quantum superposition. These phenomena mean that the electronic Hamiltonian itself contains all possible 'reaction pathways', therefore, the pure electronic dynamics analysis can be used as a useful beginning for the exploration of the complex nonadiabatic process.

The current SF dynamics are close to the earlier experimental and theoretical observations, which supports the $CT$-mediated mechanism in the pentacene dimer. For the current five-state model, one of the $CT$ states ($AC$)  plays an important role in the SF dynamics due to the lower energy, and the super-exchanging mechanism is governed by this state. The introduction of the vibrational degrees of freedom would enhance the probability towads the 'reaction channel' from $LE$ to $TT$ mediated by $CT$. However, different roles of the vibrational modes were found. For some modes with the frequencies far from the quantum resonance of the electronic transitions or with very minor vibronic couplings, they are basically the bystander modes. A few modes that show the frequency resonance with the adiabatic energy gap, which drive the ultrafast SF dynamics. The other modes add the damping effects in the SF dynamics. 

Overall, the current work provides a deep and detailed understanding of the SF process in the pentacence dimer. In addition, it also verifies the efficiency of the tensor train decomposition method in the treatment of the quantum dynamics in complex systems. This can serve as the reference for other similar systems to study and explore the dynamic behaviors. Besides, the time-resolved spectroscopy is often used in the study of SF processes. The spectral simulation is very important to provide some new perspectives to understand experimental observations. This should be an important topic in the future.

\section{Author Declarations}
\subsection{Conflict of interest}
The author has no conflicts to disclose.

\section{Data Availability}
The data and code that support the findings of this study are available from the corresponding author upon reasonable request.


\begin{acknowledgement}
\noindent
This work is supported by NSFC projects (Nos. 21933011, 22333003, 21873112 and 22173107). The authors thank the Supercomputing Center, Computer Network Information
Center, Chinese Academy of Sciences; National Supercomputing Center in Shenzhen for providing computational resources.
\end{acknowledgement}
\newpage
\setcounter{figure}{0}
\renewcommand{\thefigure}{A\arabic{figure}}
\section{Appendix}
\begin{center}
  Table 1. The parameters $V_{kl}$ in the molecular Hamiltonian (in eV). Here, the energy of $S_0S_0$ state is set to 0.000 eV, and the number in the bracket denotes its index of the state order in the calculations.
\end{center}
\begin{center}
  \begin{tabular}{c | c c  c  c  c  c }
    \hline
    &  $S_0S_0$ (0) &  $S_1S_0$ (1) &  $S_0S_1$ (2) &  $AC$ (3) &  $CA$ (4) &  $TT$ (5) \\                                                
    \hline
    $S_0S_0$ (0) &  0.000  &  0.000  &  0.000  &  0.000  &  0.000  &  0.000 \\
    $S_1S_0$ (1) &  0.000  &  2.005  & -0.006  &  0.099  &  0.079  &  0.000 \\
    $S_0S_1$ (2) &  0.000  & -0.006  &  2.010  &  0.075  &  0.086  &  0.001 \\
    $AC$ (3)     &  0.000  &  0.099  &  0.075  &  2.266  &  0.001  & -0.090 \\
    $CA$ (4)     &  0.000  &  0.079  &  0.086  &  0.001  &  3.063  &  0.064 \\
    $TT$ (5)     &  0.000  &  0.000  &  0.001  & -0.090  &  0.064  &  1.900 \\
    \hline
 \end{tabular}
\end{center}
\newpage
\begin{center}
  Table 2. The parameters of frequencies and the electronic-phonon couplings (in eV) for all $A_1$ modes.
\end{center}
\begin{center}
  \begin{tabular}{c | c | c | c | c | c }
    \hline
    frequency &  $S_1S_0$ &  $S_0S_1$ &  $AC$ &  $CA$ &  $TT$ \\                                                
    \hline
    0.032516  & -0.015686   &   0.000000   &   -0.028929  &    0.014225   &   -0.013981 \\
    0.074878  &  0.007802   &   0.000000   &    0.018566  &   -0.012140   &    0.009755 \\
    0.077694  & -0.006104   &   0.000000   &   -0.005746  &   -0.000797   &   -0.000511 \\
    0.094445  & -0.050825   &   0.000000   &   -0.028576  &   -0.016968   &   -0.043225 \\
    0.099183  & -0.007636   &   0.000000   &   -0.020361  &   -0.011362   &   -0.030902 \\
    0.128104  &  0.025348   &   0.000000   &    0.020311  &    0.022932   &    0.021080 \\
    0.144957  & -0.042922   &   0.000000   &   -0.023447  &   -0.015052   &   -0.050176 \\
    0.149243  &  0.051825   &   0.000000   &    0.032766  &    0.034232   &    0.082734 \\
    0.166037  &  0.043884   &   0.000000   &    0.023269  &    0.023148   &   -0.029459 \\
    0.179442  & -0.086685   &   0.000000   &   -0.057275  &   -0.041311   &   -0.098372 \\
    0.182880  &  0.056013   &   0.000000   &    0.031587  &    0.016630   &    0.127282 \\
    0.182919  &  0.019303   &   0.000000   &   -0.000000  &    0.000000   &   -0.000000 \\
    0.185854  &  0.024438   &   0.000000   &    0.014207  &    0.014684   &   -0.041824 \\
    0.192871  & -0.086619   &   0.000000   &   -0.018448  &   -0.061165   &   -0.146108 \\
    0.196718  &  0.027400   &   0.000000   &    0.001095  &    0.012696   &   -0.045782 \\
    0.395757  & -0.008727   &   0.000000   &   -0.013256  &   -0.025695   &   -0.016480 \\
    0.397594  & -0.008275   &   0.000000   &    0.001241  &   -0.014583   &   -0.008516 \\
    0.401401  & -0.017255   &   0.000000   &    0.000943  &   -0.021579   &   -0.017810 \\
    \hline
 \end{tabular}
\end{center}
\begin{center}
  \begin{tabular}{c | c | c | c | c | c }
    \hline
    0.032516  &  0.000000   &  -0.015686   &    0.014225  &   -0.028929   &   -0.013981 \\
    0.074878  &  0.000000   &   0.007802   &   -0.012140  &    0.018566   &    0.009755 \\
    0.077694  & -0.006104   &   0.000000   &   -0.005746  &   -0.000797   &   -0.000511 \\
    0.094445  &  0.000000   &  -0.050825   &   -0.016968  &   -0.028576   &   -0.043225 \\
    0.099183  &  0.000000   &  -0.007636   &   -0.011362  &   -0.020361   &   -0.030902 \\
    0.128104  &  0.000000   &   0.025348   &    0.022932  &    0.020311   &    0.021080 \\
    0.144957  &  0.000000   &  -0.042922   &   -0.015052  &   -0.023447   &   -0.050176 \\
    0.149243  &  0.000000   &   0.051825   &    0.034232  &    0.032766   &    0.082734 \\
    0.166037  &  0.000000   &   0.043884   &    0.023148  &    0.023269   &   -0.029459 \\
    0.179442  &  0.000000   &  -0.086685   &   -0.041311  &   -0.057275   &   -0.098372 \\
    0.182880  &  0.000000   &   0.056013   &    0.016630  &    0.031587   &    0.127282 \\
    0.182919  &  0.000000   &   0.019303   &    0.000000  &   -0.000000   &   -0.000000 \\
    0.185854  &  0.000000   &   0.024438   &    0.014684  &    0.014207   &   -0.041824 \\
    0.192871  &  0.000000   &  -0.086619   &   -0.061165  &   -0.018448   &   -0.146108 \\
    0.196718  &  0.000000   &   0.027400   &    0.012696  &    0.001095   &   -0.045782 \\
    0.395757  &  0.000000   &  -0.008727   &   -0.025695  &   -0.013256   &   -0.016480 \\
    0.397594  &  0.000000   &  -0.008275   &   -0.014583  &    0.001241   &   -0.008516 \\
    0.401401  &  0.000000   &  -0.017255   &   -0.021579  &    0.000943   &   -0.017810 \\
    \hline
 \end{tabular}
\end{center}
\newpage
\begin{center}
  \begin{figure}[htbp]
      \centering
      \includegraphics[width=16cm,height=7.60cm]{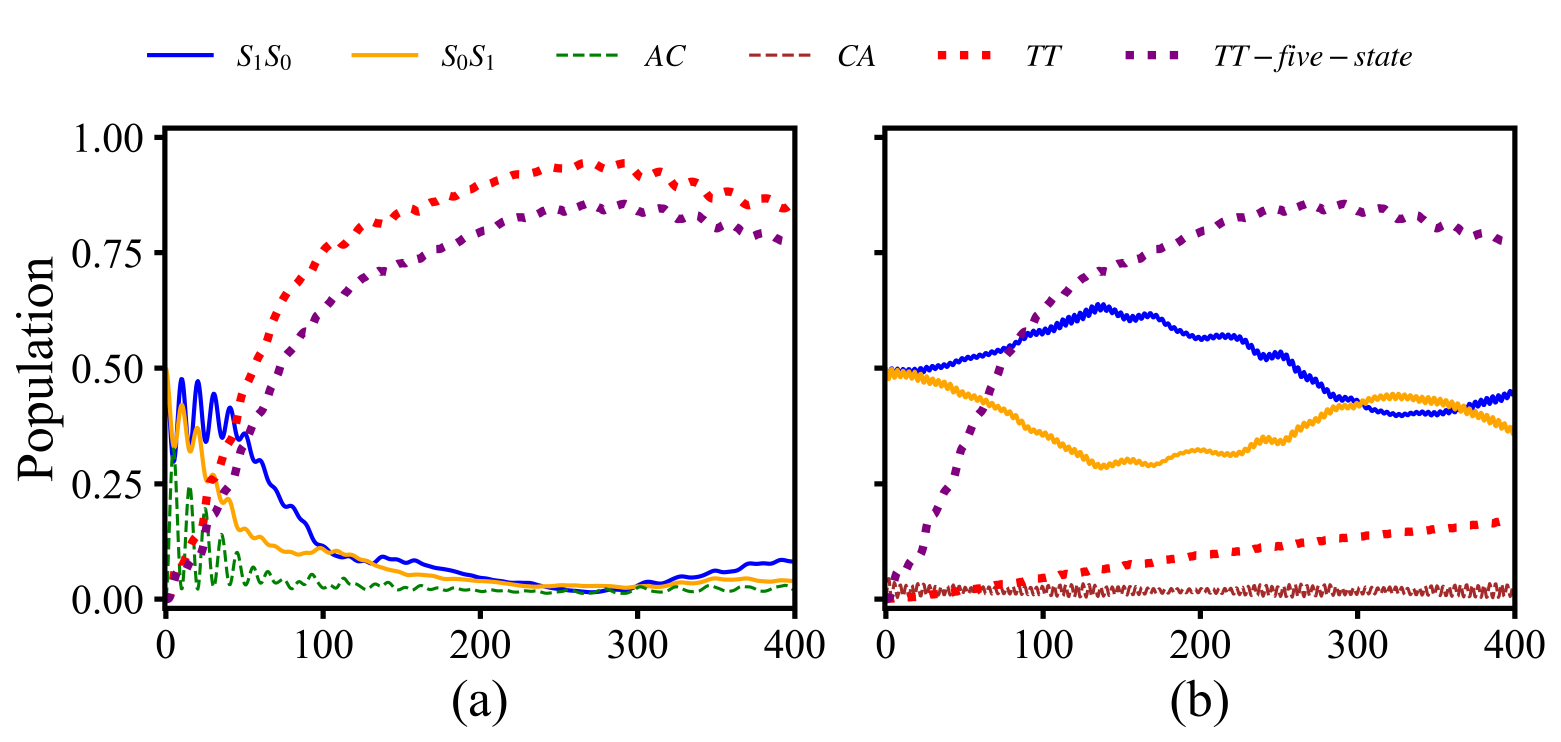}
      \caption{the time-dependent population of reduced four-state model, (a) only the charge-transfer state $AC$ was considered, (b) only the charge-transfer state $CA$ was included.}
  \end{figure}
\end{center}

\begin{center}
  \begin{figure}[htbp]
      \centering
      \includegraphics[width=12cm,height=20cm]{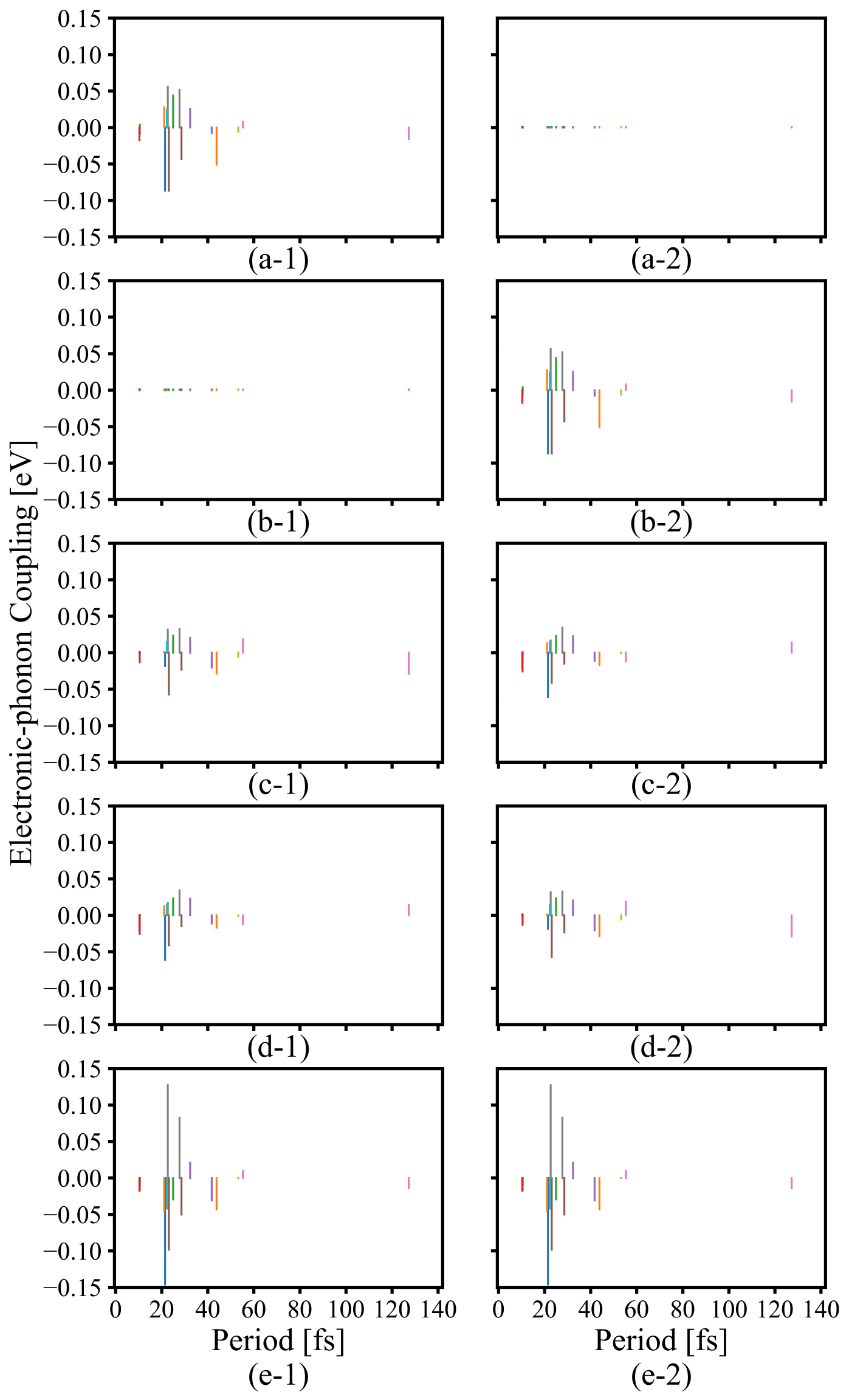}
          \caption{the electronic-phonon coupling of different reserved modes. (a-d) correspond to the modes coupled to the $S_1S_0$, $S_0S_1$, $AC$, $CA$, $TT$, respectively. 1 and 2 denote the modes related to different monomers.}
  \end{figure}
\end{center}

\newpage

\bibliography{achemso-demo}



  
  
  
  

\end{document}